\documentclass{elsart}
\usepackage{graphicx,color}
\usepackage{amsmath}

\begin{document}

\begin{frontmatter}
\title{Quarkyonic Matter and Chiral Symmetry Breaking}

\author[bnl]{Larry McLerran},
\author[wroclaw]{Krzysztof Redlich},
\author[munich]{Chihiro Sasaki}
\address[bnl]{Physics Dept. and Riken Brookhaven Research Center, 
Bdg. 510A, Brookhaven National Laboratory Upton, NY-11973, USA}
\address[wroclaw]{Institute for Theoretical Physics, 
University of Wroclaw, PL-50204, Wroclaw, Poland}
\address[munich]{Technical University, Munich D-85748, 
Garching, Germany}

\begin{abstract}
The appearance of a new phase of QCD, Quarkyonic Matter  in the limit of large number of colors
is studied within Nambu-Jona-Lassinio effective chiral model coupled to the Polyakov loop. The
interplay  of this novel QCD phase with  chiral symmetry restoration and color deconfinement is
discussed. We find that at vanishing temperature and at large $N_c$, the quarkyonic transition
occurs at densities only slightly lower than that expected for  the chiral transition. This property
is also shown to be valid at finite temperature if the temperature is less than that of  deconfinement.
The position and $N_c$-dependence of chiral critical end point is also discussed.
\end{abstract}


\end{frontmatter}

%%%%%%%%%%%%%%%%%%%%%%%%%%%%%%%%%%%%%%%%%%%%%%%%%%%%%%%%%%%%%%%%
%%%%%%%%%%%%%%%%%%%%%%%%%%%%%%%%%%%%%%%%%%%%%%%%%%%%%%%%%%%%%%%%
\section{Introduction}

The conventional view on
% folklore about
the phase diagram of QCD  is that high density strongly interacting matter   is divided into two phases:
the confined and the de-confined~\cite{itoh,carruthers,cabibbo,collins,baym,freedman,shuryak,kapusta,bielefeld,polyakov,susskind,svetitsky,kuti}.  The phase diagram of QCD  as a
function of baryon number chemical potential $\mu$ and temperature $T$ as shown in Fig. 1,
was originally envisioned by Cabibbo and Parisi~\cite{cabibbo} and has changed little conceptually since.
Until very recently, the possible new physics in the QCD phase diagram were Color Superconducting phases
which might be important at asymptotically high baryon number density and low temperatures~\cite{alford}.
In addition, a combination of  efforts of Lattice Gauge Theory (LGT)~\cite{LGT} and effective model 
calculations~\cite{pisarski,pisarski3,fukushima,weise,wambach,feinberg} have given new insight on
the position,  the order and the universal properties \cite{universal} of the QCD phase diagram.

%%%%%%%%%%%%%%%%%%%%%%%%%%%%
\begin{figure}
\begin{center}
\includegraphics*[width=8cm]{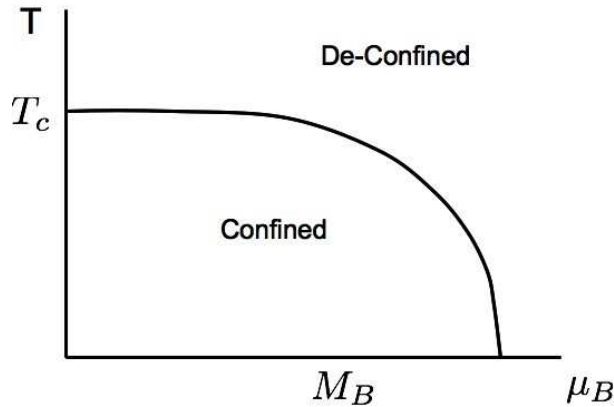}
\caption{The "phase diagram" of QCD presented as envisaged by Cabibbo and Parisi. } 
\label{cabibbo0}
\end{center}
\end{figure}
%%%%%%%%%%%%%%%%%%%%%%%%%%%%%%%%%%%%%%%%%%%%%

It has been argued recently that there may be an additional phase, the Quarkyonic Phase, of
dense QCD~\cite{pisarski1}~\footnote{
  Speculations about related phases of matter were made  in early strong coupling lattice 
  studies of QCD at high baryon number density~\cite{susskind1,kawamoto}.
}. This phase was rigorously shown to exist in the limit of a large number of colors $N_c$.  
In this limit, both the exponential of the free energy of a heavy test quark added to the system
\begin{equation}
e^{-\beta F_q} = \frac{1}{N_c} \langle L \rangle\,,
\end{equation}
and the baryon number density are order parameters~\cite{pisarski1}.  The baryon number is an order
parameter since $\langle N_B \rangle \sim e^{-\beta (M_B-\mu_B)}$.  Thus,  for temperatures of $T
\sim \Lambda_{QCD}$ and with $M_B\sim N_c$  the  $\langle N_B \rangle \sim e^{-\kappa N_c}\to 0$ at
large $N_c$, as long as the baryon number chemical potential is small compared to the baryon mass, 
i.e. $\mu \ll M_B$. When $\mu_B \ge M_B$, then baryons begin to populate the system and the baryon number 
density is non-zero.  In Ref.~\cite{pisarski1}, it was argued that there are at least three phases in QCD 
at large $N_c$: the mesonic-phase  which is confined and has zero baryon number density,
the de-confined phase which has finite baryon number density,  and the quarkyonic-phase which has
finite baryon number density and is confined.  The role of the chiral phase transition was not
established.

The reason for the existence of the quarkyonic world was because for any finite value of chemical
potential for quarks, $\mu_Q = \mu_B/N_c$, quark loops do not affect the confining potential.  The
de-confinement temperature is at some $T_c$ and is independent of $\mu_Q$.  Therefore when baryons
are added to the system, one can compress the baryons to very high chemical potential compared to 
$\Lambda_{QCD}$, and the baryonic matter remains confined.  When $\mu_Q \ge \sqrt{N_c} \Lambda_{QCD}$, 
then the effects of the quark loops are felt on the potential and there is de-confinement,
but as $N_c \rightarrow \infty$, the density at which this occurs approaches infinity.

\begin{figure}
\begin{center}
\includegraphics*[width=8cm]{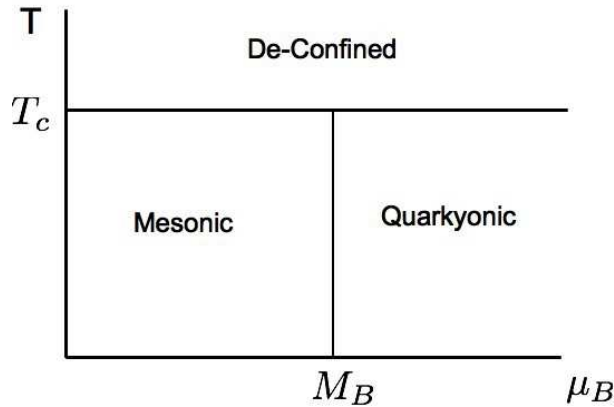}
\caption{The phase diagram of QCD in large $N_c$.  We do not display either the Chiral or Color
 Superconducting phases on this plot.}
\label{cabibbo1}
\end{center}
\end{figure}

The mesonic world is confined and has an energy density which scales as $O(1)$ in powers of $N_c$.
The de-confined  energy density scales as $N_c^2$, due to unconfined gluons.  The energy density of
the quarkyonic world scales as $N_c$, since both for baryonic matter and quark matter, the energy
density is of order $N_c$.  The quarkyonic world  may be visualized as a quasi free degenerate Fermi gas of quarks in a sea
of thermally excited mesons and glueballs.   The effects of confinement are important for quark interactions only
near the Fermi surface.
The bulk interactions deep inside the Fermi sea, even though in a confined phase, are described
by perturbation theory. The name quarkyonic was chosen since it is a combination of baryonic
and quark matter, and expresses the Yin-Yang nature of the matter.  A hypothetical phase diagram of
QCD in the large $N_c$ limit is shown in Fig.~\ref{cabibbo1}.

%%%%%%%%%%%%%%%%%%%%%%%%%%%%
%%%%%%%%%%%%%%%%%%%%%%%%%%%%%%%%%%%%%%%%%%%%%

%%%%%%%%%%%%%%%%%%%%%%%%%%%% R42 FIGURE
\begin{figure}
\begin{center}
\includegraphics*[width=8cm]{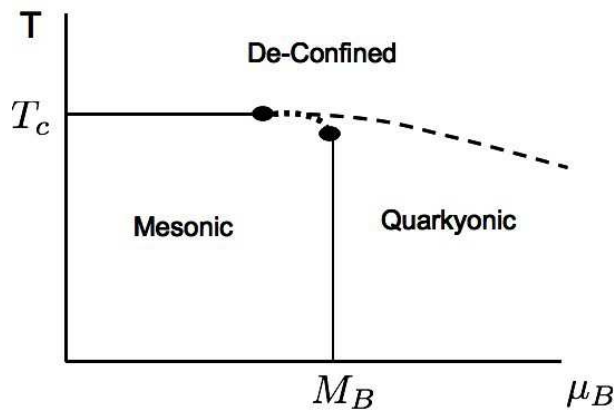}
\caption{A hypothetical phase diagram including $1/N_c$ effects (again ignoring the effects of the
Chiral transition and Color Superconductivity).} 
\label{cabibbo2}
\end{center}
\end{figure}
%%%%%%%%%%%%%%%%%%%%%%%%%%%%%%%%%%%%%%%%%%%%%

In this paper we outline a theory which allows for explicit computation in the context of the PNJL 
model of QCD~\cite{fukushima,weise}. This provides a concrete description of quarkyonic matter 
along with spontaneous chiral symmetry breaking and the effects of finite $N_c$ on the quarkyonic 
phase transition. In the large $N_c$ limit, we consider an exactly solvable model, which has the
features expected for quarkyonic matter.  We will argue that
as the baryon number density increases from zero, the confinement-deconfinement phase transition
weakens, in accord with the arguments of de Forcrand and Philipsen~\cite{deforcrand}.  We also argue as one increases
the temperature from zero, the quarkyonic phase transition weakens. The chiral phase transition, which is very close
to that of the quarkyonic phase transition is first order, and is situated almost atop the region where the quarkyonic phase transition took place, until the critical temperature is reached where it then
follows the deconfinement phase transition.

As we decrease $N_c$ from asymptotically large values, there is some point
where the first order deconfinement transition weakens and begins to disappear, as shown in Fig. 3.  Eventually the low
density confinement-deconfinement part of the phase transition completely disappears, leaving a line of first
order phase transitions and a critical point which is a remnant of the chiral-quarkyonic phase transition,
as shown in Fig. 4.  These distinct branches of the phase diagram were shown previously in the work within the PNJL model~\cite{sfr:pnjl,fuku2}, and by Miura and Ohnishi in strong coupling lattice gauge theory~\cite{onishi}.  Strictly speaking, the quarkyonic phase transition becomes a sharp cross over at finite $N_c$.  We find that in large $N_c$, the quarkyonic transition occurs at densities slightly lower than that of the chiral 
transition. 
%and at temperatures slightly above or equal to that of the confinement transition.  
This difference in density is however so small that it may be an artifact of the model.  As shown on Fig. 4, 
the cross over associated with the deconfinement phase transition continues on to higher densities, 
and becomes a separate line.  This will be explained in detail later.

%%%%%%%%%%%%%%%%%%%%%%%%%%%% R42 FIGURE
\begin{figure}
\begin{center}
\includegraphics*[width=8cm]{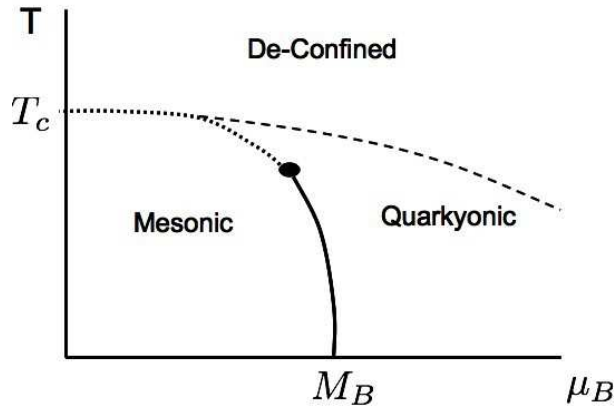}
\caption{A guess for the phase diagram of QCD for realistic value of $N_c$ (without the Chiral and
Color Superconducting phase transitions). } 
\label{cabibbo3}
\end{center}
\end{figure}
%%%%%%%%%%%%%%%%%%%%%%%%%%%%%%%%%%%%%%%%%%%%%

%%%%%%%%%%%%%%%%%%%%%%%%%%%%%%%%%%%%%%%%%%%%%%%%%%%%%%%
\section{The PNJL Model in Large $N_c$}
%%%%%%%%%%%%%%%%%%%%%%%%%%%%%%%%%%%%%%%%%%%%%%%%%%%%%%%%

In order to study the QCD phase diagram in large $N_c$ we construct a chiral model where
constituent quarks~\cite{NJL} couple to effective gluon degrees of freedom. 
Here we follow~\cite{pisarski,fukushima,weise} to introduce an extended Nambu-Jona-Lasinio model 
with Polyakov loops (PNJL model)~\footnote{
  One should keep in mind that the PNJL model describes statistical suppression of colored
  one- and two-quarks contributions which imitates color confinement.
}.

We take the Lagrangian for a constituent quark field $\psi$ as
\begin{equation}\label{lagrangian}
{\mathcal L} = \overline \psi \left( i \gamma_\mu D^\mu - m + i\mu\gamma_0 \right)\psi
{}+ {G \over 2} \left\{ (\overline \psi  \psi)^2 + (\overline \psi i\vec{\tau} \gamma_5 \psi )^2 
\right\} - U(\Phi[A], \overline \Phi [A] )\,,
\end{equation}
where $m$ is the current quark mass, $\mu$ is the quark chemical potential and $\vec{\tau}$ are
Pauli matrices. In the following we restrict our discussion to two quark flavors, $N_f=2$. 
An extension of the model to  $N_f>2$ is straightforward.

The interaction between  the quarks and the effective gluon field  is implemented through a
covariant derivative
\begin{equation}
D_\mu = \partial_\mu - iA_\mu\,, \quad \quad A_\mu = \delta_{\mu 0}A^0\,,
\end{equation}
where  $A_\mu = g A_\mu^a \frac{\lambda^a}{2}$. Here $g$ is the color SU(3) gauge coupling constant
and $\lambda^a$ are the Gell-Mann matrices. In the PNJL model, the transverse components of $A$ are
integrated out.  If we assume that these reflect short distance degrees of
freedom, the effective potential in terms of $\Phi$ and $\overline \Phi$ should be at most 6'th
order in the fields ($\Phi, \overline \Phi$), since these are the relevant operators for the three
dimensional space on which $\Phi$ and $\overline \Phi$ exist~\footnote{
  When we have a kinetic energy term $(\partial \Phi)^2$ of a scalar field $\Phi$,
  we would have to rescale $\Phi$ in such
  a way that it has the correct canonical dimension $(d-2)/2$.
  Thus, the highest order term
  with a renormalizable coupling involves the 6th order in scalar
  fields.
}. The variable $\Phi$ is defined as a trace of the Wilson line $L = P \exp\left\{ i \int_0^\beta
d\tau A_0(\vec{x}, \tau) \right\}$ in color space:
\begin{equation}
\Phi = {1 \over N_c} \mbox{Tr} L\,,
\end{equation}
and $\overline \Phi$ is the complex conjugate.

We will work in mean field approximation for the fields $\Phi$.  The potential $U$ must
respect the $Z(N_c)$ symmetry.  For large $N_c$, the most general potential is of the form~\cite{pisarski},
\begin{equation}
{U \over T^4}
=
C \left(\frac{N_c^2-1}{8}\right)
\left( -{{b_2(T)} \over 2} \overline \Phi \Phi {}+
{{b_4} \over 4} \left(\overline \Phi \Phi \right)^2
{}+ {{b_6} \over 6} \left(\overline \Phi \Phi \right)^3\right)\,,
\label{potential}
\end{equation}
with an overall constant $C$. One factorizes the $N_c^2-1$ dependence to get at high $T$ the pressure 
of an ideal gluon gas. The coefficients $b_i$ must be chosen so that at high temperatures, there is 
spontaneous breaking of the $Z(N_c)$ symmetry, and at low $T$ the symmetry is restored.
For $N_c \ge 3$, the pure gauge theory has a first order phase transition corresponding to the 
de-confinement at some temperature $T_0$. At $T=T_0$ one finds
\begin{equation}
b_2(T_0) = -\frac{3b_4^2}{16 b_6}\,.
\end{equation}
We assume that $b_2(T)$ has the following temperature dependence:
\begin{equation}
b_2(T) = a_0 + a_1\left(\frac{T_0}{T} \right) + a_2\left(\frac{T_0}{T} \right)^2\,,
\end{equation}
with constant parameters $a_i$. The Polyakov loop expectation value $\langle\Phi\rangle$ must be unity 
at asymptotically high temperature $T \to \infty$. This leads to
\begin{equation}
a_0 = b_4 + b_6\,.
\end{equation}
We fix the parameters $a_i, b_i$ and $C$ taking $b_6=1$ in such a way that (\ref{potential}) 
describes the LGT observations for $\Phi(T_0)$ and pressure $P(T) = -U(T)$ in $SU(N_c=3)$
pure gauge theory~\cite{LGT SU(3)}. We list the resulting parameters in Table~\ref{param gl}.
%%%%%%%%%%%%%%%%%%%%%%%%%%%%%%%%%%%%%%%%%
\begin{table}
\begin{center}
\begin{tabular*}{8cm}{@{\extracolsep{\fill}}cccccc}
\hline
$a_0$  & $a_1$   & $a_2$   & $b_4$   & $b_6$  & $C$\\
\hline
$0.787$ & $0.333$ & $-1.13$ & $-0.213$ & $1.00$ & $5.35$ \\
\hline
\end{tabular*}
\end{center}
\caption{Set of parameters for the Polyakov-loop effective potential. }
\label{param gl}
\end{table}
%%%%%%%%%%%%%%%%%%%%%%%%%%%%%%%%%%%%%%%%%%
We assume that $T_0$ is approximately independent of $N_c$ in $U$. This is supported by a finding in lattice QCD
for several $N_c$~\cite{lattice tc nc} where the deconfinement transition temperature approaching 
from low temperature phase is parameterized as
\begin{equation}
T_0(N_c;\mu=0) = \sqrt{\sigma}\left( 0.596 + \frac{0.453}{N_c^2}\right)\,,
\end{equation}
with the string tension $\sigma$ being independent of $N_c$. Thus at large $N_c$, 
$T_0$ is ${\mathcal O}(1)$.

In this paper we are interested in structural features of the PNJL model. Here we simply assume that there is
a first order transition in the absence of quarks, and then determine the effect of quarks on this transition.  
Note that in the large $N_c$ limit, to leading order, the quarks do not affect the effective 
potential $U$. We will therefore assume a rigid background of fields $\Phi$ and explore the 
consequences for fermion-induced phase transitions.

Note also, that we need to specify which root of $\Phi$ we take when the $Z(N_c)$ symmetry is broken. 
We take the real positive root, the root which is selected if there is a small breaking of the $Z(N_c)$ 
symmetry induced by fermions.

Using bosonization, this Lagrangian can be re-expressed as
\begin{equation}
{\mathcal L} = {}- U - {{\sigma^2 + \pi^2} \over {2G}}
{}- i \mbox{Tr}\ln S^{-1}\,.
\end{equation}
with quark propagator
\begin{equation}
S^{-1} = i \gamma_\mu \partial^\mu - \gamma_0  A^0 - M\,,
\end{equation}
and dynamical  quark mass
\begin{equation}
M = m -\left( \sigma + i \gamma_5 \vec{\tau} \cdot \vec{\pi} \right)\,.
\end{equation}
The field $\sigma$ in the mean field approximation,
\begin{equation}
\langle \sigma \rangle = G \langle \overline \psi \psi \rangle\,,
\end{equation}  
gets a nonzero expectation value from solving the gap equation.

%%%%%%%%%%%%%%%%%%%%%%%%%%%%%%%%%%%%%%%%%%%%%%%
\section{The NJL  sector}
%%%%%%%%%%%%%%%%%%%%%%%%%%%%%%%%%%%%%%%%%%%%%%

In the following we first consider the solution of the fermionic sector of  the theory at large
$N_c$ and at finite $T$ and $\mu$. There are two parameters we need  to fix in vacuum: the 3-momentum
cutoff $\Lambda$ and the 4-Fermion coupling $G$. With $f_\pi \sim 92 {\rm MeV}$ for QCD, we have the 
relationship
\begin{equation}
\frac{f_\pi^2}{N_c} = M_Q^2 \int {{d^3p} \over (2\pi)^3} {{\Theta(\Lambda - \mid \vec{p} \mid )}
\over E_p^3}
\sim   {M_Q^2 \over {2\pi^2}}\left( \ln(2\Lambda/M_Q) -1 \right)\,, \label{fpi}
\end{equation}
with the constituent quark energy $E_p = \sqrt{|\vec{p}|^2 + M_Q^2}$. On the right hand side of this
equation, we have dropped all terms which vanish in the  limit $\Lambda \rightarrow \infty$. Note
that  $f_\pi^2$ should be of order $N_c$ and $G \sim 1/N_c$ for large $N_c$. We also have
\begin{equation}
\frac{\langle \overline \psi \psi \rangle}{N_c}
=
4M_Q \int ~ {{d^3p} \over {(2\pi)^3}}
{{\Theta(\Lambda - \mid \vec{p} \mid)}
  \over E_p }
\sim
{M_Q \over \pi^2} \left\{ \Lambda^2 -M_Q^2\left(\ln(2\Lambda/M_Q)  -
   {1 \over 2} \right)\right\}\,,
\end{equation}
where the dynamical quark mass $M_Q$ is obtained as the  solution of the gap equation
\begin{equation}
M_Q^2 \left\{\ln(2\Lambda/M_Q) - {1 \over 2} \right\} = \Lambda^2 \left( 1-{2\pi^2 \over{N_c N_f
G\Lambda^2}} \right)\,. \label{gap}
\end{equation}
For $N_f = 2$ taking $m=5$ MeV with constituent quark mass  $M_Q = 320$ MeV
and $f_\pi = 92\,\mbox{MeV}\times\sqrt{3/N_c}$  one gets:  $\Lambda = 646$~MeV and $G =
10.2\,{\rm GeV}^{-2}\times(3/N_c)$.

At finite temperature and density, the thermodynamic potential becomes
\begin{equation}
\Omega(T,\mu) =   U + {{(M-m)^2} \over {2G}}
{}- {T \over 2} \sum_n \int {{d^3p} \over {(2\pi)^3}}
\mbox{Tr} \ln S^{-1}((2n+1)T, \vec{p})/T\,.
\end{equation}
The trace in the above expression can be done explicitly leading to
\begin{eqnarray} \label{eq14}
\Omega 
&=& -2N_f T \int {{d^3p} \over{(2\pi)^3}} \left[ \mbox{Tr}\ln\left\{1+Le^{-(E_p-\mu)/T}
\right\}
{}+ \left( L \rightarrow L^\dagger, \mu \rightarrow -\mu \right) \right]  \nonumber\\
&&
{}- 2N_c N_f \int {{d^3p} \over {(2\pi)^3}} E_p \Theta(\Lambda^2-|\vec{p}|^2)
{}+{(M-m)^2 \over {2G}} + U\,.
\end{eqnarray}
Taking $\Phi=1$ in Eq.(\ref{eq14}) reproduces a standard NJL potential
\begin{eqnarray}
\Omega_{\rm NJL} 
&=& -2N_c N_f T \int\frac{d^3p}{(2\pi)^3} \left[ {\ln}(1 + e^{-\beta(E_p-\mu)})
{}+ {\ln}(1 + e^{-\beta(E_p+\mu)}) \right]
\nonumber\\
&& 
{}- 2 N_c N_f \int\frac{d^3p}{(2\pi)^3} E_p\,\Theta(\Lambda^2 - |\vec{p}|^2) {}+
\frac{(M-m)^2}{2G}\,, \label{omeganjl}
\end{eqnarray}
which quantifies the thermodynamics of interacting quarks in the deconfined phase.

%%%%%%%%%%%%%%%%%%%%%%%%%%%%%%%%%%%%%%%%%%%%%%%%%%%%%%%%%%%%%%%%%%%%%%%%
\section{Solving the Fermion Sector of the  Theory at large $N_c$}
%%%%%%%%%%%%%%%%%%%%%%%%%%%%%%%%%%%%%%%%%%%%%%%%%%%%%%%%%%%%%%%%%%%%%%%%

The fermionic   contribution to the thermodynamic potential  is obtained from Eq. (\ref{eq14}) as
\begin{equation}
\delta \Omega_f
= -2 N_f T \int \frac{d^3p}{(2\pi)^3}~ \left[ \mbox{Tr}\ln \{1 + L e^{-\beta(E_p-\mu)} \}
{}+ \left( L \rightarrow L^\dagger, \mu \rightarrow -\mu \right) \right]\,. 
\label{omegaf}
\end{equation}
Suppose we are at $\mu \le M_Q$.  In this case the exponential factors inside the logarithm are all
less than one.  Moreover, $L$ is a unitary matrix, $L L^\dagger = 1$, so that we can always expand 
the logarithm as a series  in $L$ (To see this, work in a diagonal representation where each element 
of $L$ is a pure phase).  Using $\mbox{Tr}\,{\ln}A = {\ln}\,\mbox{det}A$ for a matrix $A$, one finds
\begin{eqnarray}
&& \mbox{Tr}\,{\ln}\left( 1 + L e^{-\beta(E_p-\mu)} \right)
\nonumber\\
&& = {\ln }\left[ 1 + N_c \Phi\,e^{-\beta(E_p-\mu)} {}+ F_2(\langle L^2 \rangle, \langle L
\rangle)\, e^{-2\beta(E_p-\mu)}
\right.
\nonumber\\
&& \left.
{}+ \cdots
{}+ F_p(\langle L^p \rangle, \langle L^{p-1} \rangle, \cdots, \langle L
\rangle)\,e^{-p\beta(E_p-\mu)}
{}+ \cdots + e^{-N_c\beta(E_p-\mu)} \right]\,.
\end{eqnarray}
The coefficient of $p$-quark contribution $F_p$ is a function that contains the trace
of at most $L^p$.

In the confined phase, the first nonzero contribution occurs when the determinant is expanded to
order $L^{N_c}$.  This contribution is of order $e^{-\kappa N_c}$ for temperatures of order 
$\Lambda_{\rm QCD}$ and $\mu$  a finite amount below $M$. Therefore quarks do not contribute to 
the effective potential  in the confined phase, since they are exponentially suppressed. In the 
de-confined phase, all terms of order $L^p$ contribute, so the fermion are important again.  
Note however, that to compute the contribution of the fermion determinant requires evaluating
contributions of order $\langle L^p \rangle$ which cannot be simply re-expressed in terms of
$\langle L \rangle$.  In fact to really deal with the large $N_c$ limit requires an effective
potential for the Wilson line in the pure gauge sector which retains contributions such as 
$\langle L^p \rangle$.  Nevertheless, even when in the de-confined phase, where in large $N_c$ 
there is an expectation value of $L$, it should be a good approximation to expand out the fermion 
determinant to first order in $L$.  This is known for a free fermion theory where the Boltzman 
statistics result is accurate within $20\%-30\%$. Thus,  to include the non-leading order effects 
in $N_c$, we will therefore expand the determinant to first order in $L$.  Note that in the confined 
phases of the theory, this should always be a very good approximation since there, in leading order in
$1/N_c$, $\Phi = 0$ and non-leading effects generate only a small expectation value for $\Phi$. The
terms of higher order in $L$ are expected to be further suppressed.

Consequently,  when  expanding the  fermion determinant in positive powers of $e^{-\beta (E-\mu)}$
for $\mu \le M_Q$    we conclude that whenever we are confined then baryons are exponentially 
suppressed.  In large $N_c$, there is no affect of temperature on the boundary between the 
quarkyonic and confined phase.   There is no contribution of fermions to the expectation value 
of the $\sigma$ field, so that the chiral symmetry is unaffected by an increase in either density 
or temperature while in the confined phase. There is no feed back in large $N_c$ of the gluons onto the expectation values of 
the $\sigma$ field, again in the confined phase.  We expect that the above  is changed when going  to finite $N_c$,  and that 
the expectation value of the $\sigma$ field will weaken as we go to higher temperatures.  This is 
the conventional picture that increasing temperature destabilizes the chiral condensate.  
Note also that at finite $N_c$, we should expect that the fermions will feed back upon the gluon 
potential.  Assume $N_c$ is large so that we can  expand the fermion contribution and keep only the 
first term in $\Phi$,   then this term will act to destabilize the first order deconfinement transition.
Its effect increases as the baryon number density increases.
We are therefore led to a picture like that of Philipsen and de Forcrand:  The confinement transition 
weakens with increasing baryon number density.  Therefore, if there is a critical endpoint for realistic 
$N_c$, it is the critical end point of the chiral phase transition, not that of confinement.  We will 
discuss these non-leading effects in $N_c$ in a later section.

Now let us determine the properties of the system when $\mu \ge M_Q$.  We assume we are in the
confined phase.  We have to rearrange the determinant whenever $\mu \ge E_p$. We first note that the
\begin{equation}
\mbox{Tr}\ln L  = 0\,,
\end{equation}
because it is an element of the $SU(N_c)$ group so that we can rewrite
\begin{eqnarray}
&&
\mbox{Tr}\,{\ln}\left( 1 + L e^{-\beta(E_p-\mu)} \right)
\nonumber\\
&&
= \mbox{Tr}\,{\ln}L e^{-\beta(E_p-\mu)}
\left( 1 + L^\dagger e^{-\beta(\mu-E_p)} \right)
\nonumber\\
&&
= \mbox{Tr}\,{\ln}L {}+ \mbox{Tr}\,{\ln}e^{-\beta(E_p-\mu)} {}+ \mbox{Tr}\,{\ln} \left( 1 +
L^\dagger e^{-\beta(\mu-E_p)} \right)
\nonumber\\
&&
= \mbox{Tr}\,\beta(\mu-E_p) {}+ \mbox{Tr}\,{\ln} \left( 1 + L^\dagger e^{-\beta(\mu-E_p)}
\right)\,.
\end{eqnarray}
Thus, Eq.~(\ref{omegaf}) becomes
\begin{eqnarray}
\delta \Omega_f
&=&  -2 N_f T  \int\frac{d^3p}{(2\pi)^3} \mbox{Tr}
\left[ \Theta(E_p-\mu)
\left\{ \ln(1+L e^{-\beta(E_p-\mu)})
 + \left( L, \mu \rightarrow L^\dagger, -\mu \right)\right\}
\right.
\nonumber \\
&&
\left.
{}+ \Theta (\mu-E_p) \left\{ \beta (\mu - E_p)
{}+ \ln(1+L^\dagger e^{-\beta(\mu-E_p)}) +
 \left( \mu \rightarrow -\mu \right) \right\}
 \right]\,.
\end{eqnarray}
The second term in the above equation is part of the ideal gas contribution for a zero
temperature degenerate gas of quarks.  Note that it is no longer exponentially suppressed.
Thus,  non-interacting  quarks contribute to the free energy. This contribution persists even 
in the confined phase.  In the confined phase in large $N_c$, this is the only term present, 
so it represents the contribution of quarkyonic matter.

For large but finite $N_c$  if we are at low temperatures, far away from the deconfined phase,
the expectation value of $\Phi$ is small.  Therefore, we can expand the determinant to first
order.  This shows how we generate an explicit expectation value, and should show that its effects
on the quarkyonic phase boundary are small.

So lets assume that the expectation value of $\Phi$ is very small in cold matter. Then, the vacuum
contribution plus the finite chemical potential contribution of the fermions are of the form
\begin{eqnarray}
\Omega_{\rm quark} 
& = & -2N_c N_f \int {{d^3p} \over {(2\pi)^3}} E_p \left( \Theta(\Lambda - E_p)
{}- \Theta(\mu - E_p) \right)  
\nonumber \\ 
& & - 2N_c N_f \mu \int {{d^3p} \over {(2\pi)^3}} \Theta(\mu - E_p)\,.
\end{eqnarray}

The extremization of this effective potential leads to the gap equation,
\begin{equation}
{{\pi^2} \over {GN_cN_f}} = \int_{p_F}^\Lambda ~dp~{p^2 \over {E_p}}\,,
\end{equation}
where the integral is cut off at the Fermi momentum $p_F = \sqrt{\mu^2 - M^2}$.
The chiral phase transition point is determined by the vanishing of the chiral condensate,
$M \rightarrow 0$.  Using Eq.~(\ref{gap}) and the above  gives us the solution for the position 
of the chiral phase transition as
\begin{equation}
\mu_{\rm chiral}^2(T=0) =  M_Q^2 \left\{ \ln(2\Lambda/M_Q) - {1 \over 2} \right\}\,,
\label{much zero T}
\end{equation}
with the vacuum quark mass $M_Q$.
Putting in numbers, one finds that $\mu_{\rm chiral} = 1.04\, M_Q$ in $N_c = 3$~\footnote{
  The model with present parameters shows a first order phase transition in low temperatures.
  Nevertheless, an actual numerical calculation with $m=5$ MeV gives almost identical critical 
  value for a first order transition, $\mu_{\rm chiral}=1.05\,M_Q$.
}.  Note, that this
equation tells us that the chiral transition occurs at densities slightly greater than that where
the quarkyonic transition occurs ($\mu = M_Q$). The cutoff $\Lambda$ is independent of $N_c$ for
large $N_c$ and thus the coefficient ${\ln}(2\Lambda/M_Q) - 1/2$ has only a weak $N_c$-dependence.
The significance of this conclusion is subject to many uncertainties. One cannot therefore 
with certainty conclude nor rule out, that there is an intermediate phase of unconfined constituent
quarks, as has been argued by Feinberg et al.~\cite{feinberg}.  In any case, the position of the 
chiral and the quarkyonic phase occur at numerically close, or perhaps identical, values of $\mu$.

%%%%%%%%%%%%%%%%%%%%%%%%%%%%%%%%%%%%%%%%%%%%%%%%%%%%%%%%%%%%%%%%%%%%%%%%%%%
\section{Gluon effects on the chiral phase transition in large $N_c$}
%%%%%%%%%%%%%%%%%%%%%%%%%%%%%%%%%%%%%%%%%%%%%%%%%%%%%%%%%%%%%%%%%%%%%%%%%%%%

Let us begin by considering the response of the Fermion determinant to a non-zero expectation value
of $\Phi$.  If we are at high temperature in the deconfined phase, then it is a good approximation
to set $\Phi = 1$.  In this case,  we have an ideal gas of quarks whose thermodynamics is described by
a pure NJL model given in~(\ref{omeganjl}). In the following,  we will be interested in the
restoration of chiral symmetry where near the restoration point, we may approximate the mass as
small.  In this limit, one  determines the chiral critical line from the gap equation
\begin{equation}
1 = \frac{N_c N_f G\Lambda^2}{2\pi^2}
{}- \frac{G N_c N_f}{\pi^2}
\int_0^\infty dp\,p \left( \frac{1}{1 + e^{\beta(p-\mu)}} {}+ \frac{1}{1 +
e^{\beta(p+\mu)}} \right)\,,
\end{equation}
where the integration can be  carried out analytically. Using Eq.~(\ref{gap}) for the constituent
quark mass $M_Q$ in vacuum, this equation becomes
\begin{equation}
\mu^2 + \frac{\pi^2}{3}T^2 = M_Q^2 \left\{ \ln(2\Lambda/M_Q) - {1 \over 2} \right\}\,. \label{phase
njl}
\end{equation}
If we put in numbers, we see that everywhere in the de-confined phase, in the limit $L = 1$,
 chiral symmetry is restored for $T_{\rm chiral} \sim 177$ MeV at $\mu = 0$ at $N_c = 3$.

If we include the effect of nonzero $\Phi$, by expanding the fermion determinant to first order in
$\Phi$ (an approximation which is good to 20\%-30\% even for $\Phi = 1$), one obtains
\begin{eqnarray}
&& \Omega = U + \frac{(M-m)^2}{2G}
{}- 2 N_f N_c \int\frac{d^3p}{(2\pi)^3} E_p\,\Theta(\Lambda-|\vec{p}|) {}+ \delta\Omega_f\,,
\\
&&
\delta\Omega_f = - 2 N_c N_f T \int\frac{d^3p}{(2\pi)^3}
\left[ \Theta(E_p-\mu)
\Phi \left( e^{-\beta(E_p-\mu)} + e^{-\beta(E_p+\mu)} \right)
\right.
\nonumber\\
&&
\left.
\quad\quad
{}+ \Theta(\mu-E_p)\left\{ \beta(\mu-E_p)
{}+ \Phi \left( e^{-\beta(\mu-E_p)} +
e^{-\beta(\mu+E_p)} \right) \right\} \right]\,,
\end{eqnarray}
where a difference between expectation values of $\Phi$ and $\bar{\Phi}$ at finite $\mu$ is
neglected. This leads to the gap equations for $M$ and $\Phi$,
\begin{eqnarray}
&&
\Lambda\sqrt{\Lambda^2 + M^2} - M^2\ln\left( \frac{\Lambda + \sqrt{\Lambda^2+M^2}}{M}\right)
{}- \left( 1- \frac{m}{M}\right)\frac{2\pi^2}{N_c N_f G}
\nonumber\\
&&
= 4\Phi MT\mbox{cosh}(\mu/T)K_1(M/T)
{}+ \Theta(\mu-M)\left[ p_F\mu - M^2\ln\left(\frac{p_F + \mu}{M}\right)
\right.
\nonumber\\
&&
\quad
\left.
{}- 4\Phi\int_M^\mu dE \sqrt{E^2-M^2}\,\mbox{cosh}((E-\mu)/T)
\right]\,,
\label{gapM}
\\
&&
T^3 C \left(\frac{N_c^2-1}{8}\right)\left[ -b_2(T) + b_4\Phi^2 + b_6\Phi^4 \right]\Phi
\nonumber\\
&&
= \frac{2N_c N_f}{\pi^2}\left[ \mbox{cosh}(\mu/T)M^2 T K_2(M/T)
\right.
\nonumber\\
&&
\left.
\quad
{}+ \Theta(\mu-M)\int_M^\mu dE \sqrt{E^2-M^2}\,\mbox{sinh}((E-\mu)/T)
\right]\,.
\label{gapPhi}
\end{eqnarray}

One can see that the gap equations take a simple form assuming a second order chiral transition,
\begin{eqnarray}
&&
\mu^2 + 4\Phi T^2 = M_Q^2\left[ \ln(2\Lambda/M_Q) - \frac{1}{2}\right]\,,
\label{gap1}
\\
&&
\mu^2 + 2T^2 = \frac{C \pi^2 (N_c^2-1)}{16N_c N_f}T^2
\left[ -b_2(T) + b_4\Phi^2 + b_6\Phi^4 \right]\Phi\,.
\label{gap2}
\nonumber\\
\end{eqnarray}
Thus, in the large $N_c$ limit they are two gap equations which describe the quark 
and gluon sectors without any interference. Eq.~(\ref{gap1}) coincides with Eq.~(\ref{much zero T})
and thus the chiral phase transition is indicated by a straight line $\mu_{\rm chiral}(T) = 
\mu_{\rm chiral}(T=0)$. This also dictates the order of phase transition for any $T$.
Eq.~(\ref{gap2}) determines the first-order deconfinement transition described by $\Phi(T)$
and thermodynamics now depends only on $T$.

%%%%%%%%%%%%%%%%%%%%%%%%%%%%%%%%%%%%%%%%%%
\begin{figure}
\begin{center}
\includegraphics[width=10cm]{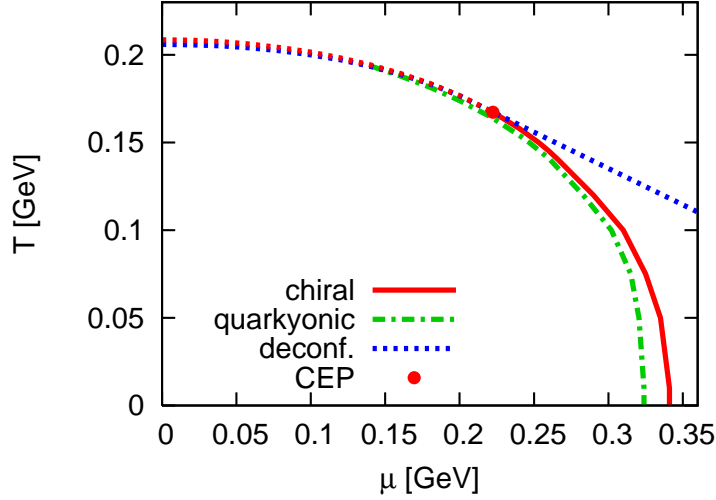}
\caption{The phase diagram for $N_c=3$ obtained in our model with the current quark mass $m=5$ MeV.
The solid lines indicate a first order phase transition while the dashed lines
cross over transitions. The critical end point (CEP) is indicated by a dot 
on the chiral phase boundary lines.} 
\label{nc3}
\end{center}
\end{figure}
%%%%%%%%%%%%%%%%%%%%%%%%%%%%%%%%%%%%%%%%%%%%%

Fig.~\ref{nc3} shows the model phase diagram for $N_c=3$. The chiral and deconfinement
cross over lines are identified as a maximum of derivatives $\partial M/\partial T$
and $\partial\Phi/\partial T$, respectively. The chiral and deconfinement lines are almost on top
in a wide range of $\mu$. Near the critical end point (CEP) the chiral and quarkyonic
transitions are strongly coupled and the CEP appears near the intersection of those
boundary lines. The chiral phase boundary is influenced the deconfinement transition, 
and may weaken the chiral transition and result in the appearance of a CEP.

%%%%%%%%%%%%%%%%%%%%%%%%%%%%%%%%%%%%%%%%%%
\begin{figure}
\begin{center}
\includegraphics[width=10cm]{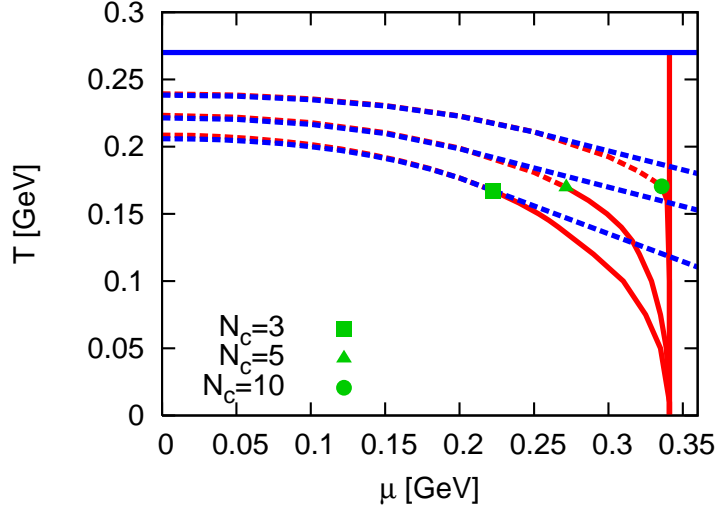}
\caption{The deconfinement and chiral phase boundary for various $N_c$.
The horizontal  line describes the deconfinement phase boundary in $N_c \to \infty$.
The vertical line indicates the chiral phase boundary in $N_c\to\infty$.
The solid line indicate a first order phase transition while the dashed lines
cross over transitions. The symbols represent the chiral CEP for corresponding $N_c$.
}
\label{chdec}
\end{center}
\end{figure}
%%%%%%%%%%%%%%%%%%%%%%%%%%%%%%%%%%%%%%%%%%%%%

The evolution of the phase boundaries with $N_c$ is shown in Fig.~\ref{chdec}.
The chiral transition lines move to larger $\mu$ and approach the vertical line.
Correspondingly, the CEP is also shifted to the right with $N_c$ and eventually
disappears in the $N_c \to \infty$ limit. The coincidence of the chiral and deconfinement
transitions is unaffected by $N_c$. Both lines are equally shifted upward approaching
the horizontal line $T_0=270$ MeV characterizing deconfinement transition temperature
in pure gauge sector.

%%%%%%%%%%%%%%%%%%%%%%%%%%%%%%%%%%%%%%%%%%
\begin{figure}
\begin{center}
\includegraphics[width=10cm]{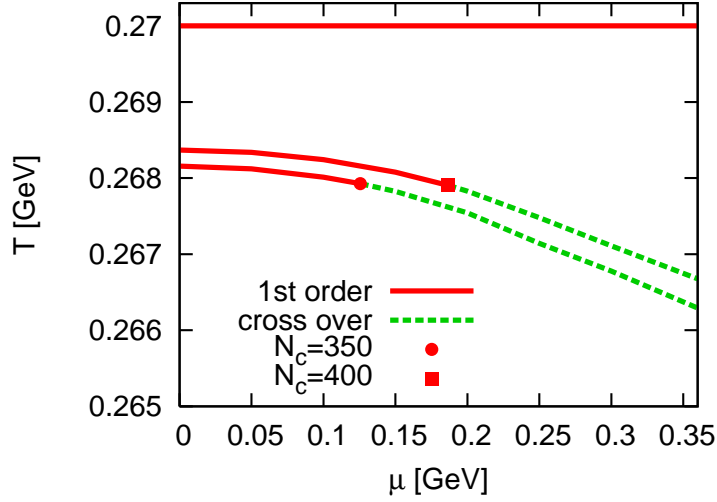}
\caption{The deconfinement transition lines for various $N_c$
The horizontal  line describes the deconfinement phase boundary in $N_c \to \infty$.
The solid lines indicate a first order phase transition while the dashed lines
cross over transitions. The symbols represent the CEP asociated with $Z(N_c)$ symmetry.}
\label{zncep}
\end{center}
\end{figure}
%%%%%%%%%%%%%%%%%%%%%%%%%%%%%%%%%%%%%%%%%%%%%

Increasing $N_c$ further, the cross over of deconfinement turns into a first order transition
and a CEP associated with the $Z(N_c)$ symmetry appears at finite $\mu$. This behavior
is indicated in Fig.~\ref{zncep}. This CEP appears as a result of quark interactions
which makes the transition weaken. Thus the CEP
disappears again in the large $N_c$ limit since quarks do not affect deconfinement.

%%%%%%%%%%%%%%%%%%%%%%%%%%%%%%%%%%%%%%%%%%
\begin{figure}
\begin{center}
\includegraphics[width=10cm]{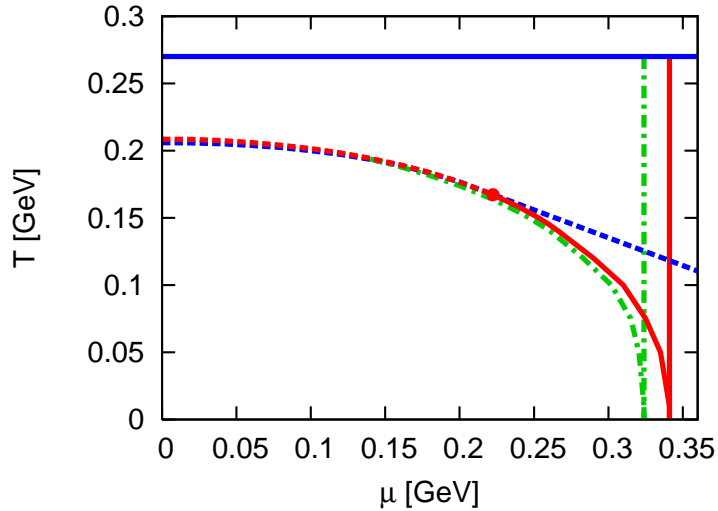}
\caption{The phase diagram of our model for $N_c=3$ and for $N_c \to \infty$.
The horizontal and vertical solid lines indicate the deconfinement and chiral
phase boundaries in $N_c \to \infty$. The vertical broken line indicates the second
order quarkyonic phase transition in $N_c \to \infty$.
}
\label{nc}
\end{center}
\end{figure}
%%%%%%%%%%%%%%%%%%%%%%%%%%%%%%%%%%%%%%%%%%%%%

Fig.~\ref{nc} shows the phase diagram for $N_c=3$ and for $N_c=\infty$. The model describes
three distinct phases in the large $N_c$ limit which agrees with the argument made in~\cite{pisarski1}.
(There may or may not be a fourth phase in a narrow range of chemical potential
where there is baryon number but chiral symmetry is not restored. However, such a phase 
might also be an artifact of the model used in our calculation.)
For realistic $N_c=3$ the order of phase transitions is changed due to the quark-gluon interference.
Nevertheless, one sees a remnant of the phase structure in large $N_c$ along with a deformation of
the boundaries including finite $N_c$ effects.

%%%%%%%%%%%%%%%%%%%%%%%%%%%%%%%%%%%%%%%%%%%%%%%%%
\section{Summary and Conclusions}
%%%%%%%%%%%%%%%%%%%%%%%%%%%%%%%%%%%%%%%%%%%%%%%%%%

In this paper, we have shown, that  the first order phase
transition and associated critical end point are driven by the
chiral  rather then  deconfinement transition.  The lines of both
these  transitions sit nearly atop the cross over associated with
the remnants of the first order quarkyonic phase transitions of
the large $N_c$ limit.
The critical end point itself may appear where the cross over from the 
quarkyonic and the deconfinement phase transitions intersect.

It is interesting that within this theory, chiral symmetry breaking may 
occur at temperatures below that of deconfinement.   This has been addressed 
in several papers, and might be associated with 
parity doubling~\cite{glozman,vento,rischke}.

Clearly, chiral symmetry restoration in QCD may be more
complicated than that which appears in PNJL type models.
The Fermi surface of the quarkyonic phase is presumably associated 
with confined particles, and it may be, that there is a chiral condensate 
associated with this Fermi surface.  Therefore, although chiral symmetry 
may be approximately restored at or very near the quarkyonic transition,
the full restoration of chiral symmetry might require much high energy density, 
and might ultimately be associated with the deconfinement transition.

We have also seen that the cross overs associated with deconfinement and 
the quarkyonic phase transition are remnants of the first order phase transition 
which occurred in the limit of large $N_c$. Thus, also at finite $N_c$, 
the lines of cross overs reflect the phase structure seen in the large $N_c$ limit. 
It is interesting, that the chiral critical end point typically appears 
at the juncture of the confinement and quarkyonic cross overs.
This property may be more general and may appear beyond  the PNJL model analysis. 
This is because, the large change of baryon number density associated with the 
quarkyonic phase transition may drive a first order chiral phase transition  
until the increase in the number of degrees of freedom associated with  deconfinement  
destabilizes it, resulting in a critical end point.

In summary, it is fair to say that the conclusions from the
PNJL model analysis are at best suggestive of what the true
structure of high baryon number density matter might be.

%%%%%%%%%%%%%%%%%%%%%%%%%%%%%%%%%%%%%
\section*{Acknowledgments}
%%%%%%%%%%%%%%%%%%%%%%%%%%%%%%%%%%%%%

C.S. acknowledges stimulating discussions with N.~Kaiser and W.~Weise.
The work of C. Sasaki has been supported in part by the DFG cluster of excellence 
``Origin and Structure of the Universe''.
K. Redlich acknowledges partial support of the Polish Ministry of Science and Higher  Education (MENiSW).
The research of L. McLerran is supported under DOE Contract No. DE-AC02-98CH10886.

%%%%%%%%%%%%%%%%%%%%%%%%%%%%%%%

\end{document}